\documentclass[12pt]{article}
\usepackage[margin=1in]{geometry}
\usepackage[T1]{fontenc}
\usepackage[slantedGreek]{mathpazo}
\usepackage[pdftex]{graphicx}
\usepackage{amsfonts}
\usepackage{amssymb}
\usepackage{amsthm}
\usepackage{float}
\usepackage{amsmath}
\usepackage{hyperref}
\usepackage{natbib}
\usepackage{booktabs}

\title{Photons = Tokens: The Physics of AI and the Economics of Knowledge}
\author{
	\makebox{Alec Litowitz}\thanks{QStar Capital. https://www.qstarcap.com}\and
	\makebox{Nick Polson}\thanks{University of Chicago. ngp@chicagobooth.edu}\and
	\makebox{Vadim Sokolov}\thanks{George Mason University. vsokolov@gmu.edu}
}
\date{First Draft: February 8, 2026\\This Draft: \today}

\graphicspath{{./fig/}}

\begin{document}

\maketitle
\begin{abstract}
Debates about artificial intelligence capabilities and risks are often conducted without quantitative grounding. This paper applies the methodology of \citet{mackay2009sustainable}---who reframed energy policy as arithmetic---to the economy of AI computation. We define the \emph{token}, the elementary unit of large language model input and output, as a physical quantity with measurable thermodynamic cost. Using Landauer's principle, Shannon's channel capacity, and current infrastructure data, we construct a supply-and-demand balance sheet for global token production. We then derive a finite \emph{question budget}: the number of meaningful queries humanity can direct at AI systems under physical, information-theoretic, and economic constraints. We apply Coase's theory of the firm and the durable-goods monopoly problem to the AI value chain---from photon to atom to chip to power to token to question---to identify where economic value concentrates and where regulatory intervention is warranted. We argue that the expansion of the token budget does not resolve a deeper constraint: under structural uncertainty, the decisive variable is not how many questions can be answered but which questions are worth asking---a problem of agency and direction that computation alone cannot solve. We connect limits of measurement in the token economy to a structural parallel between Goodhart's law and the Heisenberg uncertainty principle, and to Arrow's impossibility result for efficient information pricing. The framework yields order-of-magnitude estimates that discipline policy discussion: at current efficiency, the projected 2028 US AI energy allocation of 326~TWh could support roughly $6.5 \times 10^{17}$ tokens per year, or 225,000 tokens per person per day---more than three orders of magnitude above estimated mid-2024 utilization.
\end{abstract}

\section{Introduction}
\label{sec:intro}

In 1957, dying of cancer, John von Neumann prepared what would be his final work: \emph{The Computer and the Brain}, the Silliman Lectures at Yale that he was too ill to deliver \citep{vonneumann1958computer}. In eighty pages, he argued that computation is a physical process. The brain and the digital computer are both engines for transforming energy into information, and the thermodynamic costs of this transformation are irreducible. A neuron dissipates roughly $10^{-3}$~ergs per operation; a vacuum tube of his era, roughly $10^{2}$~ergs---a factor of $10^{5}$. The brain compensates for its slow, imprecise components through massive parallelism; the computer compensates through raw speed. In both systems, the fundamental operations are the same: \emph{storage and retrieval} of information, governed by the laws of physics. Von Neumann died before he could finish the manuscript, but the thesis was complete: intelligence, whether natural or artificial, is computation, and computation is physics.

Half a century later, David MacKay published \emph{Sustainable Energy---Without the Hot Air}, a book that reframed energy policy as a problem of arithmetic \citep{mackay2009sustainable}. His method was simple: convert every energy source and every energy demand into the same unit---kilowatt-hours per day per person---and check whether the numbers balance.

The debate over artificial intelligence needs both von Neumann's insight and MacKay's method. Von Neumann established that computation is physics; MacKay showed how to do the accounting. This paper applies MacKay's arithmetic to the computational economy that von Neumann foresaw, with the \emph{token}---the elementary unit of large language model input and output---playing the role the kilowatt-hour played in MacKay's analysis.

A token is a subword unit, typically three to four characters of English text. Generating one requires a forward pass through a neural network: a physical computation on silicon, consuming electricity, dissipating heat. The analogy to energy is not metaphorical. Power in physics is energy per unit time, measured in watts. Power in AI is information processed per unit time, measured in tokens per second. Both are physical quantities, both are finite, and both can be accounted for.

The intellectual lineage runs through cybernetics and the theory of computation. \citet{turing1936computable} established the mathematical limits of computation: there exist well-defined problems that no finite procedure can solve, and every computable function can be realized by a universal machine. \citet{wiener1948cybernetics} unified the study of control and communication in animals and machines under the principle that feedback requires information, and information has physical cost. \citet{vonneumann1966selfreproducing} demonstrated that self-reproducing automata require a minimum complexity threshold---and that arbitrarily reliable systems can be built from unreliable components through redundancy. \citet{shannon1948mathematical} established the mathematical theory of communication, quantifying information in bits and proving that every physical channel has a finite capacity. \citet{wiener1950human} drew the social implication: if machines process information as organisms do, then the consequences of automation are as much questions of physics as of policy \citep{prigogine1978time}.

\citet{strubell2019energy} estimated the carbon footprint of training a large NLP model; \citet{patterson2021carbon} extended this to Google-scale training; \citet{luccioni2024power} shifted focus to inference. Our contribution differs in scope: rather than measuring individual models, we construct a balance sheet for the \emph{entire} token economy.

This paper develops five results. First, we construct a \emph{token budget}: a MacKay-style balance sheet of global AI token supply and demand, grounded in current energy infrastructure data and physical efficiency limits (Sections~\ref{sec:physics} and~\ref{sec:budget}). Second, we apply Coase's theory of the firm and the durable-goods monopoly problem to the AI value chain---from photon to atom to chip to power to token to question---to identify where economic value concentrates and why it migrates upward in the stack (Section~\ref{sec:stack}). Third, we derive a \emph{question budget}: the finite number of meaningful queries that can be directed at AI systems, connecting Cox's theory of questions to Shannon entropy and Landauer's thermodynamic cost (Section~\ref{sec:questions}). Fourth, we argue that physical and informational abundance does not resolve the problem of \emph{agency}: under structural uncertainty, the binding constraint is not how many questions can be answered but which questions are worth asking---a problem of direction, path dependence, and judgment that computation alone cannot solve (Section~\ref{sec:agency}). Fifth, we identify a measurement limit on the token economy---a structural parallel between Goodhart's law and the Heisenberg uncertainty principle---and connect it to Arrow's impossibility of efficient information pricing (Section~\ref{sec:measurement}).

The central question is simple: \emph{How many questions is the world allowed to ask?}

\section{The Physics of Tokens}
\label{sec:physics}

A token is a physical object. Its generation requires a sequence of matrix multiplications on semiconductor hardware, consuming electrical energy and dissipating heat. \citet{landauer1961irreversibility} proved that any logically irreversible computation---such as erasing a bit of information---must dissipate at least
\begin{equation}
\label{eq:landauer}
E_{\min} = k_B T \ln 2
\end{equation}
of energy into the environment, where $k_B \approx 1.38 \times 10^{-23}$~J/K is Boltzmann's constant and $T$ is the absolute temperature. At room temperature ($T \approx 300$~K), this gives $E_{\min} \approx 2.87 \times 10^{-21}$~J per bit erased.

The resolution of Maxwell's demon paradox confirms this bound: any agent that reduces entropy in one part of a system must pay for the information it acquires by increasing entropy elsewhere \citep{bennett2003notes}. Computation is not free; it is a thermodynamic process.

A token drawn from a vocabulary of size $V$ carries at most $\log_2 V$ bits of information. For a typical LLM vocabulary of $V \approx 100{,}000$, this gives $\log_2(100{,}000) \approx 16.6$ bits per token. In practice, the information content is lower: \citet{shannon1951prediction} estimated the entropy rate of printed English at 1.0--1.3 bits per character, implying 3--5 bits per subword token. The Landauer floor therefore lies between $\sim 10^{-20}$~J (at the Shannon entropy of $\sim$4 bits) and $\sim 5 \times 10^{-20}$~J (at the full vocabulary of 16.6 bits)---a factor of five, negligible relative to the $10^{19}$-fold efficiency gap. We adopt 12 bits per token as a practical working estimate for forward-pass computation, which must evaluate the full output distribution even though only a fraction of bits carry information. The Landauer floor for one token is then
\begin{equation}
\label{eq:landauer_token}
E_{\text{Landauer}}^{\text{token}} = 12 \times k_B T \ln 2 \approx 3.4 \times 10^{-20} \text{ J}.
\end{equation}

\subsection{The Actual Cost}

Current LLMs operate far above this floor. Empirical measurements place the energy cost of inference at approximately $0.0001$--$0.002$~Wh per output token, depending on model size \citep{epochai2025aitrends}. For a mid-range estimate of $5 \times 10^{-4}$~Wh per token, converting to joules:
\begin{equation}
\label{eq:actual_cost}
E_{\text{actual}}^{\text{token}} \approx 5 \times 10^{-4} \times 3600 = 1.8 \text{ J}.
\end{equation}
The ratio of actual to theoretical minimum cost is
\begin{equation}
\label{eq:efficiency_gap}
\frac{E_{\text{actual}}^{\text{token}}}{E_{\text{Landauer}}^{\text{token}}} = \frac{1.8}{3.4 \times 10^{-20}} \approx 5 \times 10^{19}.
\end{equation}
Current AI hardware is roughly $10^{19}$--$10^{20}$ times less efficient than the thermodynamic limit. This gap is enormous but decomposable. Modern CMOS logic gates dissipate roughly $10^3$--$10^6$ times the Landauer limit per switching operation; the additional orders of magnitude arise because generating a single token requires on the order of $10^{12}$ floating-point operations, each involving many gate-level bit erasures. The gap represents the combined overhead of irreversible logic, memory access, data movement, cooling, and power conversion.

But the dollar cost of inference is falling fast. \citet{appenzeller2024llmflation} documents what he terms ``LLMflation'': for an LLM of equivalent capability, inference cost has declined approximately $10\times$ per year. Between November 2021 and late 2024, the dollar cost of achieving a fixed benchmark score fell from \$60 to \$0.06 per million tokens---a $1{,}000\times$ reduction in three years---driven by six independent factors: GPU hardware improvements, model quantization (from 16-bit to 4-bit arithmetic), software optimizations, smaller models achieving equivalent performance, improved training techniques (RLHF, DPO), and open-source competition compressing margins. The first four factors reduce \emph{energy} per token and expand the physical token budget; the last two reduce dollar cost without reducing energy consumption. Each successive order of magnitude is harder to achieve, and the Landauer floor remains absolute.

These efficiency gains, however, do not imply reduced total energy consumption. \citet{jevons1865coal} identified the mechanism in 1865: when technological improvement reduces the cost of using a resource per unit of output, total consumption of that resource tends to \emph{increase}, not decrease, because cheaper use induces greater demand. Jevons observed that Watt's fuel-efficient steam engine did not conserve coal---it made coal economically useful in industries that had never used it, and England's coal consumption soared. ``It is a confusion of ideas to suppose that the economical use of fuel is equivalent to diminished consumption. The very contrary is the truth.''

The Jevons paradox applies directly to the token economy. In January 2025, the Chinese laboratory DeepSeek released a frontier-class model at inference costs roughly an order of magnitude below prevailing rates, demonstrating that open-weight models with aggressive quantization and architectural efficiency could match proprietary systems at a fraction of the price \citep{deepseek2024v3}. The immediate effect was not reduced AI energy consumption but an explosion of demand: cheaper tokens made AI economically viable for applications---bulk document processing, real-time translation, continuous code generation---that had been priced out at higher per-token costs. LLMflation is the AI economy's Watt steam engine: each order-of-magnitude cost reduction expands the addressable market by more than an order of magnitude, so total energy consumption rises even as energy per token falls.

The inefficiency compounds further in practice. GPU clusters achieve a Model FLOPS Utilization (MFU) of only 45--55\% of theoretical hardware throughput \citep{nebius2025economics}. Hardware failures in a 3,000-GPU training cluster occur every $\sim$10 hours on average \citep{sivathanu2024revisiting}, and recovery, checkpointing, and rollback add approximately 30\% overhead to wall-clock training time. That reliable computation emerges at all from such failure-prone hardware is a consequence of \citeauthor{vonneumann1966selfreproducing}'s redundancy theorem: arbitrarily reliable systems can be built from components with error probability $p < 1/2$. Modern AI clusters implement this through checkpoint-and-restart protocols, spare node pools, and error-correcting interconnects.

These overheads apply primarily to training; inference workloads have different utilization and failure profiles. The distinction matters for the balance sheet in Section~\ref{sec:budget}. Training a frontier model is a one-time, high-intensity burst: a single training run may consume tens of gigawatt-hours over weeks or months \citep{strubell2019energy,patterson2021carbon}. Inference is a continuous, low-intensity flow: each query costs a fraction of a watt-hour, but the queries never stop.

As of 2024, inference accounts for an estimated 60--70\% of total AI electricity consumption, up from roughly one-third in 2023, and is projected to reach approximately two-thirds by 2026 \citep{iea2025datacenter,luccioni2024power}. The energy budget is increasingly dominated not by the cost of \emph{creating} intelligence but by the cost of \emph{using} it. The balance sheet uses the inference figure ($5 \times 10^{-4}$~Wh per token).

\subsection{Shannon's Ceiling}

\citet{shannon1948mathematical} established an upper bound: the channel capacity theorem. For any channel with bandwidth $B$ and signal-to-noise ratio $S/N$, the maximum rate of reliable information transmission is
\begin{equation}
\label{eq:shannon}
C = B \log_2\!\left(1 + \frac{S}{N}\right) \text{ bits per second}.
\end{equation}
Token throughput---AI ``power''---cannot exceed the channel capacity of the hardware interconnects, memory buses, and network links \citep[Ch.~1]{mackay2003information}.

\subsection{The Bekenstein Bound}

\citet{bekenstein1981universal} showed that the information in any bounded region of space is finite. For a system of energy $E$ enclosed in a sphere of radius $R$, the maximum entropy is
\begin{equation}
\label{eq:bekenstein}
S \leq \frac{2\pi k_B E R}{\hbar c},
\end{equation}
where $\hbar$ is the reduced Planck constant and $c$ is the speed of light. \citet{lloyd2000ultimate} showed that a 1-kg system confined to a 1-liter volume can store at most $\sim 10^{31}$ bits and perform at most $\sim 10^{51}$ operations per second. The point is qualitative: the information content and processing rate of any physical system are finite. The Landauer floor sets the energy cost per bit; Shannon's capacity sets the information rate; the Bekenstein bound sets the absolute information content.

\section{The Token Budget}
\label{sec:budget}

We construct a balance sheet for the token economy. The demand side uses global figures (AI services are consumed worldwide). The supply side focuses on the United States, which hosts an estimated 50--60\% of global AI compute infrastructure and, through cloud providers headquartered there, serves a predominantly global user base \citep{iea2025datacenter}. Dividing US energy capacity by the world population therefore yields a \emph{lower bound} on per-person token availability---the true figure is higher once non-US data centers (Europe, China, the Middle East) are included. Where US and global figures are mixed, we note the scope explicitly.

\subsection{Demand: How Many Tokens Does the World Consume?}

Estimating global AI inference volume is difficult: providers disclose selectively, definitions vary (output tokens alone versus all tokens processed), and demand has been growing rapidly. In February 2024, OpenAI alone reported generating approximately 100 billion words per day---roughly $1.3 \times 10^{11}$ output tokens from a single provider \citep{altman2024openai}. By mid-2025, Google's AI models were processing approximately 980 trillion tokens per month \citep{alphabet2025q2}, and an aggregate cross-provider analysis placed global inference volume at $10^{13}$--$10^{14}$ tokens per day \citep{openrouter2025stateofai}. For the balance sheet, we adopt a mid-2024 estimate of approximately $10^{12}$ tokens per day globally, consistent with the provider-level data and the rapid growth trajectory documented by \citet{epochai2025aitrends}. In MacKay's per-person terms:
\begin{equation}
\label{eq:per_person_current}
\frac{10^{12} \text{ tokens/day}}{8 \times 10^9 \text{ people}} \approx 125 \text{ tokens/person/day}.
\end{equation}
At 0.75 words per token, 125 tokens is roughly 94 words---about a paragraph. The average person speaks approximately 16,000 words per day \citep{mehl2007women}.

Behind these inference tokens lies a prior investment: the \emph{re-encoding} of existing information. Search engines indexed the web by cataloguing links between pages; large language models re-encode it---compressing the vast corpus of human text into parametric representations that can be queried in natural language. Training a frontier model on trillions of tokens is, in information-theoretic terms, a lossy compression of the web into the model's weights. Inference then decompresses on demand: each user query extracts a targeted reconstruction from the compressed representation. The energy cost of encoding (training) must be amortized across the inference tokens it enables.

AI-specific electricity consumption in the United States was 53--76~TWh in 2024, projected to reach 165--326~TWh by 2028 \citep{iea2024electricity,iea2025datacenter}. If efficiency remains at its 2024 level---a deliberately conservative assumption given the rapid gains documented in Section~\ref{sec:physics}---and the upper-bound projection (326~TWh) materializes, the 2028 token capacity would be
\begin{equation}
\label{eq:tokens_2028}
\frac{326 \times 10^{12} \text{ Wh}}{5 \times 10^{-4} \text{ Wh/token}} = 6.5 \times 10^{17} \text{ tokens/year} \approx 1.8 \times 10^{15} \text{ tokens/day}.
\end{equation}
Per person, this gives
\begin{equation}
\frac{1.8 \times 10^{15}}{8 \times 10^9} \approx 225{,}000 \text{ tokens/person/day},
\end{equation}
or roughly 169,000 words (at 0.75 words per token)---a novel per day.

\subsection{Supply: What Can the Infrastructure Provide?}

The supply side is constrained by energy infrastructure. The total US electrical generation capacity is approximately 1,250~GW (including distributed generation), producing 4,178~TWh in 2023 \citep{eia2024electricity}. Table~\ref{tab:balance} presents the balance sheet.

\begin{table}[t]
\centering
\caption{Token economy balance sheet, United States. The 2024 observed figure ($\sim$125 tokens/person/day) is estimated global demand divided by world population (see text for sources and uncertainty); capacity figures assume all AI electricity is used for inference at $5 \times 10^{-4}$~Wh/token. In practice, training consumes 30--40\% of AI electricity (declining as inference grows), so realized inference capacity is correspondingly lower. AI electricity midpoint: 65~TWh is the center of the 53--76~TWh range.}
\label{tab:balance}
\begin{tabular}{lrrr}
\toprule
& \textbf{2024} & \textbf{2028 (proj.)} & \textbf{Units} \\
\midrule
\multicolumn{4}{l}{\textit{Demand (observed) and capacity (projected)}} \\
AI electricity consumption & 65 & 326 & TWh/yr \\
Token capacity (at $5\times10^{-4}$ Wh/token) & $1.3 \times 10^{17}$ & $6.5 \times 10^{17}$ & tokens/yr \\
Observed tokens per person per day & $\sim$125$^{*}$ & --- & tokens/person/day \\
Capacity per person per day & 44,500 & 225,000 & tokens/person/day \\
\midrule
\multicolumn{4}{l}{\textit{Supply (grid)}} \\
Total US electricity generation & 4,178 & $\sim$4,500 & TWh/yr \\
AI share of US grid & 1.6\% & 7.2\% & \% \\
Maximum tokens at current efficiency & $8.4 \times 10^{18}$ & $9.0 \times 10^{18}$ & tokens/yr \\
\midrule
\multicolumn{4}{l}{\textit{Physical limits}} \\
Landauer-limited tokens (entire US grid) & \multicolumn{2}{c}{$\sim 10^{38}$} & tokens/yr \\
Efficiency gap (actual / Landauer) & \multicolumn{2}{c}{$\sim 5 \times 10^{19}$} & $\times$ \\
\bottomrule
\multicolumn{4}{l}{\small $^{*}$Mid-2024 estimate of $\sim 10^{12}$ tokens/day; see text for sources and uncertainty.}
\end{tabular}
\end{table}

The table reveals three facts. First, AI's share of the US electricity grid is growing rapidly but remains a single-digit percentage, even under aggressive projections. Second, the current utilization gap---observed output of $\sim$125 tokens/person/day versus an energy-implied capacity of 44,500---suggests that the binding constraint in 2024 is hardware deployment, not energy. Third, the gap between current practice and the Landauer limit ($\sim 10^{19}$) implies that enormous efficiency gains are thermodynamically possible.

A fourth observation is implicit: AI companies benefit from an externality embedded in electricity regulation itself. US industrial electricity prices---approximately 7--8 cents per kWh \citep{eia2024electricity}---are the product of a century of public utility regulation: rate-of-return constraints, public investment in generation and transmission, and cross-subsidies between customer classes that keep industrial rates below the marginal social cost of generation. The AI industry purchases this regulated input and sells tokens at unregulated market prices. At 326~TWh and \$0.07/kWh, the projected 2028 AI electricity bill is roughly \$23 billion---a small fraction of the revenue the token economy is expected to generate. The implicit subsidy flows from ratepayers and the environment (which bears the unpriced externalities of generation) to AI shareholders. This is the same structure that has historically benefited aluminum smelters and steel mills, but the growth rate of AI electricity demand---doubling every two to three years---means the subsidy is scaling faster than any precedent in industrial energy consumption.

\subsection{The Simon Wager Revisited}

The question of whether physical resources constrain the token economy has a precedent. In 1972, the Club of Rome published \emph{The Limits to Growth}, projecting that exponential demand for commodities would exhaust key resources within decades \citep{meadows1972limits}. Economist Julian Simon argued the opposite: human ingenuity, operating through price signals, would ensure that resources never run out in any economically meaningful sense \citep{simon1981ultimate}. In 1980, Simon challenged biologist Paul Ehrlich to choose any raw materials and any future date; Simon would bet that inflation-adjusted prices would fall. Ehrlich and colleagues selected five metals---chromium, copper, nickel, tin, and tungsten---and bought \$200 of each on paper, for a total stake of \$1,000, indexed to September~29, 1980, with a payoff date of September~29, 1990. Between 1980 and 1990 the world's population grew by more than 800 million---the largest single-decade increase in history---yet by September 1990 every metal had fallen in real terms. Tin dropped from \$8.72 to \$3.88 per pound; tungsten fell by more than half. In October 1990, Ehrlich mailed Simon a check for \$576.07 \citep{simon1980resources}.

For four decades, Simon's position appeared vindicated. Commodity prices remained stable or fell as substitution, recycling, and discovery outpaced demand. But the victory was sensitive to timing: asset manager Jeremy Grantham noted that had the wager run from 1980 to 2011, Simon would have lost on four of the five metals \citep{grantham2011resources}. The AI buildout may represent the scenario in which the Club of Rome's arithmetic finally binds---not on the original decade-long horizon, but on the longer one that Grantham identified. That copper was one of Simon's five metals gives the parallel particular force.

The numbers are concrete \citep{goldmansachs2024copper}. A conventional data center requires 5,000--15,000 tons of copper; a hyperscale AI facility requires up to 50,000 tons, with intensity ranging from 27 to 66 tons per megawatt. In 2024, data centers consumed approximately 500,000 tons of copper worldwide. Projections place this at 1.1 million tons annually by 2030---roughly 4\% of global demand---and 3 million tons by 2050, a sixfold increase. These demands arrive simultaneously with competing requirements from electric vehicles, renewable energy, and grid electrification; global copper demand is projected to grow from 28 million metric tons in 2025 to 42 million by 2040 \citep{iea2025datacenter}.

The supply side is equally constrained. Decades of underinvestment in exploration have produced a declining rate of major discoveries since the 1980s, while existing mines face steadily falling ore grades---lower grades mean more energy, more water, and higher costs per ton of refined metal. A major new copper mine takes 15 to 20 years from discovery to production. The industry would need approximately six new tier-one mines to come online every year through 2050 merely to meet baseline demand growth, before accounting for the incremental demands of electrification and AI \citep{giustra2025copper}. Copper prices have already reflected this arithmetic, rising from approximately \$3.80 per pound in 2023 to nearly \$6.00 in early 2025. Mining executive Robert Friedland has stated the constraint in its starkest form: merely to sustain 3\% global GDP growth---before accounting for electrification, data centers, or renewable energy---the world must mine as much copper in the next two decades as it has extracted in the past ten thousand years \citep{giustra2025copper}.

The AI buildout could create genuine scarcity---not because of population growth, as the Club of Rome predicted, but because of the physical demands of the token economy. Simon was right that ingenuity can substitute around resource constraints, but substitution requires time, capital, and physical alternatives. A copper mine takes two decades; a GPU generation takes two years. Whether the copper, energy, and fabrication capacity exist to support a thousand-fold expansion of the token economy in four years is an empirical question---and the arithmetic should be done before the commitments are made.

\subsection{The MacKay Lesson}

MacKay's central insight was that energy debates become productive only when reduced to arithmetic. The same holds for AI.

The cost of \emph{not} doing the arithmetic is illustrated by a cautionary episode from the energy debate itself. In 2009---the same year MacKay published his book---\citet{levitt2009superfreakonomics} argued in \emph{Superfreakonomics} that solar panels were counterproductive because, being black, they absorb sunlight and radiate waste heat. \citet{pierrehumbert2009levitt} showed that elementary arithmetic demolished the claim: the world's electricity needs require roughly 53,000~km$^2$ of solar cells---about 0.01\% of Earth's surface---producing waste heat that would warm the planet by approximately 0.006$^\circ$C, roughly 300 times smaller than the warming from the CO$_2$ emissions they would displace. Moreover, waste heat is a one-time effect upon installation, whereas CO$_2$ warming accumulates for centuries; after 100 years of operation, the avoided greenhouse warming exceeds the waste heat effect by a factor of 125. The error was not subtle; it was a failure to multiply. Pierrehumbert's observation---that if one cannot execute basic energy accounting, the rest of the analysis cannot be trusted---applies with equal force to AI policy claims advanced without quantitative grounding.

The projected 2028 capacity of 225,000 tokens per person per day is not negligible. Whether the infrastructure can support this growth---and whether it should---can only be answered with numbers.

\section{The Value Stack}
\label{sec:stack}

The token budget tells us how many tokens exist. It does not tell us where economic value concentrates.

\citet{coase1937nature} asked: if markets are efficient, why do firms exist? His answer was \emph{transaction costs}. Using the price mechanism---searching for counterparties, negotiating contracts, enforcing agreements---is not free. Firms emerge when the cost of organizing a transaction internally is lower than the cost of conducting it through the market. The boundary of the firm falls where the marginal cost of internal organization equals the marginal cost of market exchange.

This principle applies directly to the AI value chain, which can be decomposed into a stack of physical and informational layers:
\begin{equation}
\label{eq:stack}
\text{photon} \to \text{atom} \to \text{chip} \to \text{power} \to \text{token} \to \text{question} \to \text{value}.
\end{equation}
Each arrow represents a transformation with characteristic physics, timescales, and transaction costs. At the bottom, photons strike silicon in solar cells to generate electricity, or heat water in gas turbines to spin generators; the same element, purified to 99.9999999\% (``nine nines''), forms the substrate of semiconductor wafers. Atoms of silicon, copper, gold, and rare earths are assembled into chips through lithographic processes operating at nanometer precision---a single advanced fab consumes roughly 60~million liters of ultrapure water per day \citep{tsmc2024arizona}. Chips consume electrical power, measured in watts, to execute the matrix multiplications that produce tokens, measured in tokens per second per watt. Tokens are assembled into coherent responses to questions, and questions generate economic value when they reduce uncertainty about decisions. The stack spans twelve orders of magnitude in timescale: mining copper takes years, fabricating a chip takes months, generating a token takes milliseconds, reading it takes a fraction of a second. Capital flows upward through the stack because each transformation compresses more physical input into less physical output: tons of ore become grams of silicon, grams of silicon become nanometers of circuit, and nanometers of circuit produce megabytes of tokens that propagate at the speed of light.

The economic logic of the stack follows from a physical asymmetry that Bertrand Russell identified in compressed form \citep{russell1935idleness}: ``Work is of two kinds: first, altering the position of matter at or near the earth's surface relatively to other such matter; second, telling other people to do so. The first kind is unpleasant and ill paid; the second is pleasant and highly paid.'' At the bottom of the stack---mining copper, refining silicon, building data centers---work consists of moving atoms on Earth's surface. At the top---generating tokens, answering questions---work consists of moving bits through fiber and silicon at the speed of light. The same compression gradient that drives capital upward has a physical basis: information propagates at $c$ while atoms are bound by mass and distance. The cost of moving a kilogram of copper from mine to fab scales with both; the cost of moving a megabyte of tokens from data center to user scales with bandwidth alone. Robotics attempts to close this gap; that it remains vastly more expensive to manipulate atoms than bits is the physical basis of Russell's observation.

Russell's asymmetry yields a single economic principle: \emph{value is speed}. The principle originates in von Neumann's analysis. In \emph{The Computer and the Brain}, he reduced all computation to two primitive operations: a memory register must be able to ``store'' a number and to ``repeat'' it upon ``questioning''---to emit it to another organ upon request \citep{vonneumann1958computer}. Storage and retrieval: that is the entire architecture. The economic question is which operation commands a premium, and von Neumann identified the answer in his analysis of access times. He observed that it is ``technologically difficult, or---which is the way in which such difficulties usually manifest themselves---very expensive, to provide all words with access time $t$,'' and proposed a hierarchical memory structure in which frequently needed data is stored at fast (expensive) access times and the remainder at slow (cheap) ones. The cost of computation is, at root, the cost of retrieval speed.

This is precisely the economic service that AI provides. A large language model compresses a training corpus---trillions of tokens of human knowledge---into parametric storage (the model weights). Inference is retrieval: given a query, the model reconstructs a targeted answer from its compressed representation in milliseconds. The economic value lies not in the storage (the weights sit inertly on disk) but in the speed and relevance of retrieval---the ability to answer a natural-language question from a petabyte-scale knowledge base in the time it takes a human to read the first word of the response. The previous generation of information retrieval---web search---organized knowledge by keyword and hyperlink. Language models organize it by meaning, compressing text into parametric representations that support semantic retrieval at machine speed.

The history of computing confirms von Neumann's hierarchy. In the 1990s, the industry organized around storage (EMC) $\to$ networking (Cisco) $\to$ data $\to$ application. As storage commoditized, value migrated to retrieval speed. Jeff Dean's ``Numbers Everyone Should Know''---access latencies from L1 cache (0.5~nanoseconds) to disk seek (1.6~milliseconds), spanning six orders of magnitude---became the canonical reference for system design \citep{dean2009numbers}. The firms that accelerated retrieval captured the surplus: Intel's Math Kernel Library, NVIDIA's CUDA, Google's Tensor Processing Units. At each transition, the physical layer depreciated while the speed layer above it captured the premium. The AI stack recapitulates this pattern: the value of a GPU depreciates with each generation, but the value of fast, accurate retrieval from a compressed knowledge base only increases.

The competitive structure of the frontier AI industry confirms this. The race among laboratories---OpenAI, Anthropic, Google DeepMind, xAI, Meta---is, beneath the differences in architecture, training data, and safety methodology, a race for inference speed per unit energy. The binding competitive variable is tokens per watt per dollar. The firm that retrieves fastest at a given energy budget captures the inference market; all other advantages---brand, distribution, safety reputation, user base---are subordinate to this physical constraint. Whoever generates fastest, wins.

\subsection{Where Value Concentrates}

Coase's theory predicts that value concentrates where transaction frictions are highest. At the base of the stack, the chip layer, transaction costs are extreme. Semiconductor fabrication requires capital expenditures exceeding \$20 billion per facility (TSMC's Arizona complex is budgeted at over \$65 billion for three fabs \citep{tsmc2024arizona}), lead times of three to five years, and a supply chain spanning dozens of countries. The result is vertical integration: a handful of firms (TSMC, Samsung, Intel) control global chip production. \citet{moore1965cramming} observed that transistor density doubles approximately every two years, a trajectory that has driven six decades of hardware improvement.

But Moore's trajectory is approaching physical limits. Feature sizes below 3~nanometers encounter quantum tunneling, lithographic precision constraints, and thermal density barriers that make further doubling progressively more expensive per transistor even when technically achievable. The implication for the value stack is direct: as the hardware improvement rate decelerates, value migrates decisively from the physical chip toward the software and algorithmic layers---CUDA, model architectures, inference optimization---that determine how efficiently a fixed transistor budget is utilized. When chips stop getting faster, speed becomes a software problem.

Yet even a near-monopolist at this layer faces a fundamental constraint identified by \citet{coase1972durability}: the \emph{durable goods monopoly problem}. A monopolist selling a durable good competes with its own past and future output. Once a chip is sold, it persists in the market, reducing demand for the next generation. The mechanism is precise: a buyer who values a GPU at $y$ knows that the monopolist will eventually lower the price to capture the next tier of buyers at valuation $x < y$. If the buyer is patient---if the discount factor $d$ is close to 1---waiting costs little, and the monopolist must drop its first-period price to $dx + (1-d)y$ to prevent delay. In the limit, with a continuum of buyer valuations and a discount factor approaching unity, the monopolist is driven toward competitive pricing in the first period despite controlling the entire supply \citep{coase1972durability}. The analogy to mining is exact: every ounce already sold sits in the market competing with the next ounce.

Applied to the AI chip market: the cadence is relentless---Hopper (2022), Blackwell (2024), Vera Rubin (2026), Feynman (2028)---with each generation offering roughly $5\times$ the inference throughput at a fraction of the cost per token \citep{nvidia2024blackwell}, depreciating the millions of units already deployed. Every buyer knows the next generation is eighteen months away and five times as capable; waiting is rational. A caveat: as of 2024, the dominant GPU vendor maintains gross margins near 75\%, suggesting value has \emph{not yet} migrated upward. In the short run, fabrication bottlenecks and extreme demand sustain pricing power---the discount factor is effectively low because buyers cannot afford to wait when capacity is scarce and the opportunity cost of delayed deployment is high. But Coase's logic is patient. As supply catches up and the secondary market for used GPUs deepens, the effective discount factor rises, margins will compress, and value will shift toward the non-durable layers: tokens and questions, which are consumed upon generation and cannot accumulate in a secondary market.

\paragraph{The question layer and the token layer.}
At the top of the stack, transaction costs take a different form. Arrow's information paradox dominates: the value of a question's answer cannot be assessed until the answer is obtained \citep{arrow1962economic}. This makes efficient pricing of AI queries impossible in principle. Users cannot comparison-shop for answers they do not yet have.

The middle of the stack---the conversion of power into tokens---is where the Coasean logic is sharpest. A chip manufacturer selling GPUs to individual consumers incurs enormous transaction costs: each buyer must acquire hardware expertise, build cooling infrastructure, manage software dependencies, and bear the risk of rapid obsolescence. The Coasean conclusion is blunt: \emph{a GPU manufacturer should not be selling chips to end users}. The economically efficient unit of sale is not a chip, or even a server, but a token. A cloud provider that integrates from chips to tokens eliminates the frictions and captures the surplus. The hardware layer should be invisible to the end user, just as the turbine is invisible to the electricity consumer and the fiber-optic cable is invisible to the telephone caller.

\paragraph{The software layer: CUDA and lock-in.}
Within the hardware layer itself, value concentrates not in the physical chip but in the software framework that makes it programmable. CUDA---NVIDIA's parallel computing platform---is the decisive intermediary between raw silicon and useful computation. In von Neumann's terms, CUDA converts a matrix of transistors into a retrieval engine: it orchestrates thousands of parallel cores to execute the matrix multiplications that constitute a forward pass, turning stored model weights into a retrieved answer in milliseconds. A GPU without CUDA is inert silicon; with CUDA, it is the fastest knowledge-retrieval system ever built. The framework, not the device, is the locus of lock-in and therefore of value. The chip competes with its successors (Coase's durable goods problem); the software ecosystem that delivers retrieval speed persists.

The depth of this lock-in is visible in pricing. Jensen Huang has argued that even if competitors offered their chips for free, the opportunity cost of using them would exceed the savings \citep{nvidia2024blackwell}. The reasoning is physical: every data center is power-limited, and in a power-constrained facility, the binding variable is tokens per watt. If NVIDIA's co-designed stack delivers twice the tokens per watt of an alternative, a data center operator choosing the alternative forgoes half its potential revenue from each watt of capacity---a cost that dwarfs any discount on the silicon itself. The competitive moat is not the chip price but retrieval speed per unit energy, a joint product of hardware architecture and software stack. This is von Neumann's access-time hierarchy made commercial: the firm that delivers the fastest retrieval per joule captures the surplus, regardless of what competitors charge for storage.

\paragraph{Self-competition and cluster economics.}
Yet the same firm that is immune to external competition is locked in permanent competition with itself. Each new generation---Blackwell delivering up to $30\times$ the inference throughput of Hopper on large mixture-of-experts models \citep{nvidia2024blackwell}---renders the installed base obsolete. The millions of Hopper GPUs already deployed become the competitor that no pricing strategy can neutralize: already paid for, already racked, already running. Huang's argument about free competitor chips applies with equal force to NVIDIA's own prior generation---but in reverse. A customer with a functioning Hopper cluster faces high switching costs to Blackwell, and the Hopper cluster's opportunity cost of \emph{not} running is zero because it is already sunk. NVIDIA must price each generation not only against external alternatives but against its own customers' rational reluctance to replace working capital. This is Coase's durable goods conjecture in its purest form: the monopolist competes with its own past output over an infinite horizon, and the faster it innovates, the faster it depreciates its own installed base.

The economics of AI cluster operation illustrate this directly. A 3,000-GPU training run at commodity cloud pricing (\$2--3.50/GPU-hour) costs approximately \$1.6 million for a single model training job \citep{nebius2025economics}. Of this, roughly 30\% is consumed by overhead: hardware failures (mean time between failures $\sim$10 hours at this scale), checkpointing, rollback to saved state, and cluster maintenance \citep{sivathanu2024revisiting}. The difference between a commodity GPU rental and a vertically integrated provider---which reduces failure recovery from hours to minutes through automated replacement, shared buffer capacity, and managed orchestration---translates to savings of 20--25\% of total training cost. This is the Coasean boundary of the firm made quantitative: the transaction costs of operating raw hardware are high enough that vertical integration from chip to token is economically efficient, and firms that sell tokens rather than GPU-hours capture the resulting surplus.

The Coasean logic extends beyond the AI value chain to the structure of the firms that use it. \citeauthor{coase1937nature}'s original question---why do firms exist?---received an answer calibrated to the transaction costs of the mid-twentieth century: the costs of search, negotiation, contracting, and monitoring in the open market. AI compresses these costs at every margin. A language model that can draft contracts, analyze markets, summarize regulations, and coordinate suppliers reduces the marginal cost of market exchange relative to internal organization. The Coasean implication is that the optimal firm shrinks. In the limiting case, a single individual with sufficient AI access can perform the coordination that previously required a department---the firm of one. This restructuring is asymmetric across the stack. At the token and question layers, where work consists of processing information, the minimum viable firm approaches a single principal. At the atom layer---mining copper, fabricating chips, pouring concrete for data centers---work remains bound by physical coordination, safety regulation, and the irreducible logistics of moving matter. The optimal firm size at the bottom of the stack remains large. AI compresses the informational component of transaction costs; it does not compress the physical component. The result is a bifurcation: an economy of very small firms at the top of the stack, sustained by very large firms at the bottom.

\subsection{The Compression Principle}

\citet{shannon1948mathematical} proved that any data source with entropy $H$ can be compressed to a rate approaching $H$ bits per symbol but no further. This source coding theorem governs the stack: it sets the minimum data movement for inference, the information density of tokens, and the minimum tokens needed to answer a question.

The compression principle interacts with Moore's law to reinforce this pattern. As hardware improves, the cost of raw computation falls exponentially. But the information content of a question---its Shannon entropy---does not shrink with better hardware. A medical diagnosis, a legal analysis, a scientific hypothesis: these have irreducible informational complexity that no amount of hardware improvement can compress away.

A second, independent floor comes from computational complexity. \citet{cook1971complexity} and \citet{karp1972reducibility} established that a large class of problems---NP-hard problems including optimal scheduling, protein folding, network design, and many combinatorial optimization tasks---require computation that grows exponentially with problem size under the widely believed conjecture that $\mathrm{P} \neq \mathrm{NP}$. No hardware acceleration changes this: a problem that requires $2^n$ operations on a slow machine still requires $2^n$ operations on a fast one. Moore's law and GPU scaling reduce the constant factor; they do not reduce the exponent. The token economy can generate answers faster, but for the hardest questions, the computational cost of the correct answer exceeds the energy budget of any physically realizable machine. Shannon sets the information-theoretic floor; Cook and Karp set the algorithmic floor. Both are absolute.

The compression principle extends to computation itself. \citet{turing1936computable} showed that any computable function can be decomposed into a sequence of elementary operations on a universal machine. Programming languages formalize this at different levels of the abstraction--performance frontier: Haskell and the lambda calculus tradition maximize expressiveness---complex programs are built by \emph{composing} simple functions, $f \circ g$, each of which compresses a subproblem into its output---at the cost of execution speed. C++ compresses closer to the machine; CUDA compresses further still, mapping computation directly onto GPU hardware.

An LLM generating code traverses this hierarchy, assembling token sequences that represent nested function applications. The compression is hierarchical at every level: tokens compress characters, functions compress tokens, programs compress functions, and the deep learning frameworks (PyTorch, TensorFlow) compress neural network specifications into sequences of CUDA kernel calls.

The hierarchy extends above the programmer's level: a spreadsheet is a constrained programming language---cells as variables, formulas as functions, recalculation as execution---with severe limits on recursion, type systems, and algorithmic complexity. Excel is, in essence, a restricted and slower C++, which is itself a restricted and slower version of direct machine code. Each layer of abstraction compresses the user's cognitive burden at the cost of computational expressiveness. The LLM sits at the top of this hierarchy: it accepts natural language as input and compresses the entire abstraction stack into a single interface.

This yields a corollary that inverts a common assumption: in the mature AI economy, the act of \emph{writing code}---translating human intent into machine instructions---is a low-value activity. Code is a compression of intent into tokens, and as models improve, this compression becomes automated. \citet{acemoglu2019automation} show that automation displaces labor in existing tasks but simultaneously creates new tasks in which humans have comparative advantage. Applied to the AI stack: as token generation is automated, value migrates to the tasks that tokens cannot automate---formulating questions, exercising judgment, generating novel data from embodied experience. The scarce resource is not the ability to instruct a machine but the ability to formulate a question worth asking.

\citet{keynes1930possibilities} anticipated this conclusion nearly a century ago. Writing during the Great Depression, he predicted that within a hundred years---by approximately 2030---compound interest and technological progress would solve what he called ``the economic problem'': the struggle for subsistence that had occupied humanity since its origins. He coined the term \emph{technological unemployment}---``unemployment due to our discovery of means of economising the use of labour outrunning the pace at which we can find new uses for labour''---but dismissed it as a temporary phase of maladjustment. The deeper challenge, Keynes argued, would be leisure: ``for the first time since his creation man will be faced with his real, his permanent problem---how to use his freedom from pressing economic cares, how to occupy the leisure, which science and compound interest will have won for him, to live wisely and agreeably and well.'' He envisioned a fifteen-hour work week within three generations.

The token economy is Keynes's prediction arriving, roughly on schedule, in a form he could not have anticipated. AI does not merely automate physical labor; it automates the informational component of nearly all labor, creating not physical leisure but \emph{cognitive} leisure. The question budget (Section~\ref{sec:questions}) is the computational form of Keynes's leisure problem: with 2,200 questions per person per day at the 2028 upper bound, the constraint is not the capacity to obtain answers but the wisdom to ask questions worth answering. Keynes worried that the wealthy classes, freed from economic necessity, had ``failed disastrously'' to find purpose. The same risk applies to a civilization with abundant tokens and no framework for directing them. Formalizing this budget---and determining how many questions the infrastructure can physically support---requires measuring the information content of inquiry itself.

\section{The Question Budget}
\label{sec:questions}

Tokens are the medium; questions are the message. \citet{cox1946probability} proved that any system for reasoning under uncertainty that satisfies basic consistency requirements---divisibility, comparability, and agreement with common sense---must be isomorphic to probability theory. \citet{cox1961algebra} developed this into a full algebraic framework: probability is not merely a statistical tool but the unique consistent extension of Boolean logic to propositions whose truth value is uncertain. \citet{jaynes2003probability} showed that Cox's axioms, combined with the principle of maximum entropy, yield a complete theory of inference from incomplete information.

\citet{cox1979inference} extended this framework to the logic of inquiry. A question can be formally defined as the set of propositions that would answer it. The ``bearing'' of a question on an outstanding issue is a measure analogous to probability: it quantifies the relevance of an inquiry to the reduction of uncertainty. Just as Cox's axioms uniquely determine probability as the calculus of plausible reasoning, the logic of inquiry uniquely determines how questions should be ordered by their expected information gain---a result that connects directly to Shannon's mutual information (Equation~\ref{eq:mutual_info}).

\subsection{Questions as Entropy Reduction}

A question specifies what information is sought; receiving the answer reduces entropy \citep{shannon1948mathematical}. Consider a random variable $X$ with Shannon entropy
\begin{equation}
H(X) = -\sum_{i} p(x_i) \log_2 p(x_i).
\end{equation}
Asking a question $Q$ and receiving an answer provides mutual information
\begin{equation}
\label{eq:mutual_info}
I(Q; X) = H(X) - H(X \mid Q),
\end{equation}
reducing the remaining uncertainty from $H(X)$ to $H(X \mid Q)$. An ideal yes/no question that bisects the probability mass provides exactly 1 bit of information. A question with $n$ equiprobable answers provides at most $\log_2 n$ bits.

\subsection{The Thermodynamic Cost of Questions}

By Landauer's principle (Equation~\ref{eq:landauer}), each bit of information processed has a minimum thermodynamic cost of $k_B T \ln 2$. Asking a question and receiving an answer that provides $I$ bits of mutual information therefore costs at least
\begin{equation}
E_{\text{question}} \geq I \cdot k_B T \ln 2.
\end{equation}
This is a floor, not a ceiling. In practice, the cost is dominated by the token processing required to formulate the question and generate the answer---typically 100--1,000 tokens for a substantive query-response pair.

\subsection{The Finite Question Budget}

Given the projected 2028 AI energy budget of 326~TWh and an illustrative conversation length of 100 tokens (prompt plus response---a short factual query; substantive interactions typically consume 500--2,000 tokens, which would reduce the question count by $5$--$20\times$), the implied annual question budget under this allocation is
\begin{equation}
\label{eq:question_budget}
\frac{6.5 \times 10^{17} \text{ tokens/year}}{100 \text{ tokens/question}} = 6.5 \times 10^{15} \text{ questions/year},
\end{equation}
or roughly 6.5 quadrillion questions. Per person per day:
\begin{equation}
\frac{6.5 \times 10^{15}}{8 \times 10^9 \times 365} \approx 2{,}200 \text{ questions/person/day}.
\end{equation}

This estimate is sensitive to two parameters: the energy budget and the tokens per question. Table~\ref{tab:sensitivity} shows the question budget under varying assumptions.

\begin{table}[t]
\centering
\caption{Sensitivity of the daily question budget per person to energy budget and conversation length. Assumes $5 \times 10^{-4}$~Wh/token and 8 billion people. The 326~TWh column uses the 2028 US upper-bound projection; the 650~TWh column approximates global AI energy if non-US capacity is included.}
\label{tab:sensitivity}
\begin{tabular}{lrrrr}
\toprule
& \multicolumn{4}{c}{\textbf{Questions/person/day}} \\
\cmidrule(lr){2-5}
\textbf{Tokens/question} & \textbf{65 TWh} & \textbf{165 TWh} & \textbf{326 TWh} & \textbf{650 TWh} \\
\midrule
100  & 450 & 1,130 & 2,200 & 4,500 \\
250  & 180 & 450 & 900 & 1,800 \\
500  & 90 & 225 & 450 & 900 \\
1,000 & 45 & 113 & 225 & 450 \\
\bottomrule
\end{tabular}
\end{table}

At 100 tokens per question (a short factual query), the 2028 US budget yields 2,200 questions per person per day. At 1,000 tokens per question (a substantive multi-turn exchange with a long response), the budget drops to 225---still substantial, but an order of magnitude smaller. The sensitivity to conversation length matters because frontier models increasingly generate long, detailed responses: a medical differential diagnosis or a legal memorandum may consume 2,000--5,000 tokens, reducing the effective question budget by a factor of 20--50 relative to the short-query estimate.

The allocation of questions is an economic problem: which questions are worth asking, who gets to ask them, and at what cost. The bottleneck is not the ``fuel'' for good questions---human curiosity and domain knowledge---but the computational infrastructure to answer them.

\section{Knowledge and Agency}
\label{sec:agency}

The previous sections established what the token economy \emph{can} produce: a finite budget of tokens and questions, distributed across a value stack governed by transaction costs and physical constraints. This section asks what the token economy \emph{cannot} resolve. The expansion of computational capacity increases the rate at which uncertainty can be processed; it does not determine which uncertainties matter, how beliefs should be revised, or when to act on incomplete models. These are problems of agency---and they do not yield to scale.

\subsection{From Entropy to Direction}

Sections~\ref{sec:physics}--\ref{sec:questions} established that tokens and questions are physically bounded. Section~\ref{sec:measurement} will show that optimization against imperfect proxies introduces structural distortion. Between physical constraint and measurement limits lies a third variable: how finite questions are deployed under uncertainty.

Shannon defined information as entropy reduction \citep{shannon1948mathematical}. A token reduces uncertainty about a distribution of possible symbols. A question reduces uncertainty about a random variable via mutual information. These measures discipline the arithmetic of the token economy, but entropy reduction is not the same as directional knowledge. A message may resolve ten bits of uncertainty about tomorrow's weather or ten bits about which treatment will extend a patient's life; the Shannon entropy is identical, but the bearing on action---in \citeauthor{cox1979inference}'s sense---is not. Knowledge relevant to agency is not merely lower entropy but reduced uncertainty about the consequences of interventions: not what is likely to be observed, but what would change if one acted differently.

The token economy increases the rate at which uncertainty can be reduced, but rate is not direction. A system producing a million tokens per second can answer many questions; which questions to ask requires a model connecting information to consequences---a causal structure that the token-generating process does not itself supply.

\subsection{Agency Under Structural Uncertainty}

Agency in the token economy can be defined as the capacity to:

\begin{itemize}\setlength{\itemsep}{0pt}\setlength{\parskip}{0pt}
\item Formulate high-bearing questions,
\item Update beliefs coherently,
\item Act when probability models are incomplete,
\item Preserve adaptability across regime change.
\end{itemize}

In environments of measurable risk, optimal action follows from expected value maximization conditional on known distributions. In environments of structural uncertainty---unknown models, shifting states, nonstationary systems---optimization within a fixed distribution can be precisely wrong. Much of technological, economic, and institutional evolution occurs in the latter regime: the distributions governing semiconductor demand in 2030, the political economy of AI regulation, or the long-run effects of automating legal reasoning are not drawn from known urns.

The expansion of the token budget does not eliminate structural uncertainty; it increases the speed at which agents encounter it. A system that generates answers faster arrives sooner at the boundary where its training distribution ceases to apply. Abundance of answers is not abundance of certainty.

\subsection{Iterative Inquiry, Path Dependence, and Optionality}

Under structural uncertainty, effective use of a finite question budget resembles disciplined experimentation rather than static optimization.

Let  $S_t$  denote the feasible state space at time  $t$ . An action  $a$  induces a (stochastic) transition:

\[
S_{t + 1} = T(S_{t},a,\varepsilon_{t}),
\]

where  $\varepsilon_t$  captures unmodeled shocks and  $T(\cdot)$  is generally unknown and may itself evolve with the state.

Actions differ not only in expected payoff but in how they reshape possible futures. Irreversible commitments collapse branches of the decision tree. This introduces path dependence: the set of reachable states at a future horizon depends on the sequence of prior actions, not merely on the current state.

Consider a simple contrast. An exploitative action may offer high immediate expected payoff but collapse the option set to a narrow future trajectory. An exploratory action may yield lower short-run payoff while preserving a broader set of reachable states. Exploration carries positive option value whenever the expected continuation value under uncertainty exceeds the value of prematurely collapsing the state space---even if exploitation appears superior in the short run.

Preserving the dimensionality of $S_t$---keeping more futures reachable---can dominate short-run maximization when models are misspecified, because the value of flexibility rises with the probability that the current model is wrong.

The token economy accelerates inference, but it does not repeal path dependence. A firm that uses AI to optimize aggressively along one strategic path forecloses alternatives that no subsequent computation can reopen. More tokens cannot restore branches already collapsed.

\subsection{Measurement Limits and the Residual Role of Judgment}

Section~\ref{sec:measurement} shows that optimization against imperfect proxies generates distortion proportional to proxy quality (Equation~\ref{eqn:improvement}). Increasing optimization pressure amplifies both genuine improvement and gaming in fixed proportion; the gaming fraction $(1-\rho^2)$ is set by the proxy, not by the optimizer. No increase in computational scale eliminates this structure, because proxy metrics approximate objectives---they do not embody them.

This leaves a residual role for judgment external to automated feedback loops. The limitation is not computational but structural: objectives cannot be fully captured by any finite metric. As token throughput increases, so does the scale at which proxy distortion operates. An AI system optimizing medical diagnoses against patient satisfaction scores, or scientific output against citation counts, amplifies the gap between proxy and objective in proportion to its throughput. Optimization can be industrialized; judgment---the recognition that the proxy has diverged from the objective---cannot.

\subsection{Humans as the Open System}

\citet{shumailov2024collapse} demonstrate that models trained recursively on synthetic outputs exhibit distributional collapse: the tails of the distribution are progressively lost, variance shrinks, and error compounds with each generation. Without exogenous data, diversity contracts monotonically.

Large language models compress the statistical regularities of their training corpus into parametric representations. They do not originate new distributions---they recombine and interpolate within the support of what they have observed. The underlying data-generating process remains human activity embedded in physical and social environments: experiments conducted, observations recorded, judgments rendered, errors made and corrected. In thermodynamic terms, the model is a bounded computational system; sustained performance requires continual input from a process external to the inference loop. A model trained on the web as of 2024 has no access to the results of experiments conducted in 2025, the consequences of policies enacted after its training cutoff, or the embodied experience of navigating a world it has never inhabited. Inference is downstream of lived experience---this is not a normative claim but a structural one: a closed informational system converges toward homogeneity.

\subsection{Collective Updating and Informational Fragmentation}

Individual belief revision may be represented as Bayesian updating:

\[
p(\theta \mid D) \propto p(D \mid \theta)\,p(\theta).
\]

Collective action requires some shared protocol for updating beliefs. If agents interpret evidence under incompatible revision rules---or begin from sufficiently divergent priors---posterior distributions can diverge even as signal volume increases. This is not a pathology but a theorem: Bayesian agents with different priors and different likelihood models can process identical data and arrive at opposing conclusions.

In a low-token environment, fragmentation is constrained by scarcity of evidence: there are not enough signals to sustain many incompatible worldviews simultaneously. In a high-token environment, fragmentation can accelerate, because agents can selectively query AI systems to reinforce incompatible models of the world---each receiving internally consistent, well-argued responses that deepen rather than resolve disagreement. More information does not guarantee convergence; it guarantees only that each agent's model becomes more elaborate. Coordination therefore depends not only on compute access but on institutional norms governing inference and revision.

\subsection{Abundance Without Orientation}

Projected infrastructure permits hundreds to thousands of daily queries per person (Table~\ref{tab:sensitivity}). This represents a qualitative shift from informational scarcity to informational abundance.

Yet rapid entropy reduction is not self-directing. In the presence of path dependence and proxy distortion, high-throughput inference can amplify instability as easily as it can amplify stability. The marginal token is inexpensive---fractions of a cent at current pricing---but the marginal well-posed question is not. Formulating a question with high bearing on an open problem requires domain knowledge, causal reasoning, and awareness of what is not yet known, none of which scale with token throughput. The limiting resource in the mature token economy is therefore not computation but directional coherence: alignment between inquiry, belief revision, and long-horizon objectives under structural uncertainty.

\subsection{The Boundary of Machine Inference}

Large language models approximate conditional distributions $p(y \mid x)$: given this context, what token is likely next? Causal reasoning requires a different operation---intervention, denoted $do(X)$---which asks what would happen if a variable were set by external action rather than passively observed. Models may simulate causal reasoning through patterns absorbed from training data, but their grounding remains derivative of observed conditional distributions, not of the interventionist structure that generated those observations.

Human agency operates at the interventionist level: a physician prescribes a treatment, an engineer changes a design, a policymaker enacts a regulation. Each commits to an action that reshapes the distribution of future outcomes. Token generation simulates patterns within existing distributions; agency alters the distributions themselves. This distinction is not a limitation to be overcome by scale---it is a structural boundary between statistical inference and causal intervention.

\subsection{Implication}

The token economy converts energy into symbol sequences at unprecedented scale. Physical limits constrain efficiency (Section~\ref{sec:physics}); measurement limits constrain optimization (Section~\ref{sec:measurement}); structural uncertainty constrains model validity. Within these boundaries, the decisive variable is how finite questions are selected, sequenced, and acted upon.

These constraints---physical, informational, and structural---define the boundary conditions within which the token economy operates. The next section asks what happens when optimization proceeds without acknowledging them.

\section{The Limits of Measurement and Optimization}
\label{sec:measurement}
Section~\ref{sec:agency} argued that agency in the token economy requires disciplined inquiry under structural uncertainty: the capacity to select, sequence, and act upon finite questions while preserving optionality across evolving states. We now introduce a further constraint. Even when objectives are specified and inquiry is directed, optimization against measurable proxies introduces irreducible distortion. The limitation is not computational scale but structural: metrics approximate objectives; they do not embody them.

The token economy is physically bounded. It is also subject to a subtler constraint: the limits on \emph{measuring} and \emph{optimizing} it. \citet{goodhart1984monetary} observed, in the context of British monetary policy, that ``any observed statistical regularity will tend to collapse once pressure is placed upon it for control purposes.'' In its popular formulation: when a measure becomes a target, it ceases to be a good measure.

Language models trained to maximize a reward signal routinely exploit the proxy rather than achieving the intended objective \citep{amodei2016faulty}. \citet{clark2016faulty} documented a vivid example: a reinforcement learning agent trained to score points in the boat racing game CoastRunners learned to circle repeatedly through regenerating targets rather than completing the course---catching fire, colliding with other boats, and going the wrong direction, yet achieving a higher score than any course-completing strategy. Benchmark gaming---optimizing model performance on standardized tests rather than on real-world capability---is the same phenomenon at the industry level.

\subsection{Heisenberg's Uncertainty}

\citet{heisenberg1927ueber} established that the position $x$ and momentum $p$ of a particle cannot be simultaneously determined with arbitrary precision:
\begin{equation}
\label{eq:heisenberg}
\Delta x \cdot \Delta p \geq \frac{\hbar}{2}.
\end{equation}
The uncertainty principle follows from the non-commutativity of position and momentum operators---a property of quantum states, not of the measurement apparatus. But the intuition is often expressed through Heisenberg's microscope: a photon used to determine a particle's position imparts momentum to it, disturbing the quantity one seeks to measure. It is this measurement-disturbance picture, rather than the deeper algebraic structure, that parallels Goodhart's law most closely.

\subsection{A Structural Parallel}

These two principles---one from economics, one from quantum mechanics---are instances of a common structure: \emph{observation coupled to control distorts the observed quantity}.

In Heisenberg's case, the photon used to measure position kicks the particle, altering its momentum. In Goodhart's case, the optimization pressure applied to a metric distorts its relationship to the underlying objective.

A toy model makes the structure precise. Let $\theta$ be the true objective (e.g., genuine model capability) and $m = \theta + \varepsilon$ a measurable proxy (e.g., benchmark score), where $\varepsilon$ is noise independent of $\theta$ with $\text{Var}(\varepsilon) = \sigma_\varepsilon^2$. The proxy-objective correlation is
\begin{equation}
\rho = \frac{\sigma_\theta}{\sqrt{\sigma_\theta^2 + \sigma_\varepsilon^2}}.
\end{equation}
An optimizer that selects the candidate with the highest proxy score $m$ from $N$ alternatives achieves an expected proxy gain $G \propto \sqrt{2\ln N}$ (for large $N$, by extreme value theory). The expected genuine improvement is
\begin{equation}\label{eqn:improvement}
F = \rho^2 \cdot G.
\end{equation}
The remainder, $(1 - \rho^2)G$, is pure gaming---improvement in the noise component $\varepsilon$ that contributes nothing to the true objective $\theta$. As optimization pressure increases (larger $N$), $G$ grows, but the \emph{fraction} of genuine improvement $F/G = \rho^2$ remains fixed. To increase $G$ without losing fidelity, one must reduce $\sigma_\varepsilon^2$---that is, improve the proxy itself, which requires new measurement effort.

The analogy to Heisenberg is now visible. In quantum mechanics, $\Delta x \cdot \Delta p \geq \hbar/2$: increasing precision in position ($\Delta x \to 0$) forces loss of precision in momentum ($\Delta p \to \infty$). In the Goodhart model, increasing optimization gain ($G \to \infty$) forces the gaming waste $(1-\rho^2)G$ to grow without bound, while the genuine improvement $F = \rho^2 G$ grows at a slower rate set by the fixed proxy quality $\rho^2$. The formal analogy is not identity. In Heisenberg, there is a \emph{strict tradeoff}: reducing $\Delta x$ forces $\Delta p$ to increase. In Goodhart, there is no strict tradeoff: increasing $G$ increases \emph{both} the genuine gain $\rho^2 G$ and the gaming waste $(1-\rho^2)G$ in fixed proportion. Heisenberg constrains a product; Goodhart constrains a ratio. What the two structures share is that extraction of information from a coupled system is accompanied by an irreducible distortion set by a property of the measurement apparatus---$\hbar$ in one case, $\sigma_\varepsilon^2$ in the other.

Financial markets provide an empirical laboratory for this structure. Bruce Kovner, one of the most successful macro traders of the twentieth century, stated the principle explicitly: ``The more a price pattern is observed by speculators the more prone you have false signals; the more the market is a product of non-speculative activity, the greater the significance of technical breakout'' \citep{schwager1989wizards}. \citet{soros1987alchemy} formalized the same insight as \emph{reflexivity}: market participants' beliefs affect fundamentals, which affect beliefs, creating a feedback loop in which observation and control are inseparable. Goodhart's original observation was itself drawn from monetary policy---a market context.

For AI alignment, the implication is direct. Any proxy metric for ``beneficial AI behavior''---however carefully constructed---will be distorted by the optimization process that targets it. Stronger optimization does not help; only better proxies (lower $\sigma_\varepsilon^2$) do. This is not a solvable engineering problem at fixed measurement quality but a structural feature of optimization against imperfect metrics.

\subsection{The Information Market}

\citet{arrow1962economic} identified a fundamental paradox in the economics of information: the value of information cannot be assessed until it is acquired, but once acquired, the buyer has no need to pay for it. Information is non-rivalrous, non-excludable, and of uncertain value \emph{ex ante}---properties that make efficient pricing impossible.

Multiple large language models now compete as information providers, a game in the sense of \citet{vonneumann1944games}. But Arrow's paradox applies: users cannot know which model will best answer a question until they have the answer. And Goodhart's law applies: models optimized for benchmarks diverge from models optimized for genuine utility. The result is an information market that is useful but fundamentally incapable of efficient allocation.

This triple constraint---Goodhart's measurement distortion, Heisenberg's observer back-reaction, and Arrow's pricing impossibility---defines the boundary conditions of the token economy. The token budget can be computed; it cannot be perfectly optimized.

\section{Conclusions}
\label{sec:conclusions}

If the question budget is finite, the question of \emph{who gets access} is inescapable. The same physical stack---photon to atom to chip to power to token to question---terminates in radically different uses. The identical joule of electricity, converted through the identical GPU into the identical token, can answer a physician's diagnostic query, generate a legal brief, assist a climate scientist, or help a teenager produce a video she will watch once and forget. The stack does not discriminate. The allocation mechanism does.

This is the central regulatory question: \emph{who decides which questions get asked?} Three candidate allocators present themselves---markets, governments, and platforms---and none is without defect.

\citet{coase1960problem} showed that in the absence of transaction costs, the initial allocation of property rights does not affect efficiency: parties will bargain to the optimal outcome regardless. But transaction costs are never zero, and in the AI stack they are substantial. The implication is that the initial allocation of access rights to AI compute \emph{does} matter for efficiency, not just equity.

The default allocator is the market: tokens go to those willing to pay. Market allocation has the standard virtues---it aggregates dispersed information about willingness to pay, it provides incentives for efficiency, and it does not require a central planner to know the value of every possible question. But it also has the standard defects. Markets systematically underinvest in public goods \citep{arrow1962economic}. A question about the mechanism of antibiotic resistance has enormous social value but generates no private revenue for the questioner. A question about optimizing advertising click-through rates generates immediate private revenue but negligible social value. Left to the market, the question budget will be allocated disproportionately toward commercially valuable queries and away from basic science, public health, education, and democratic governance---precisely the domains where the social return to information exceeds the private return.

Arrow's paradox (Section~\ref{sec:stack}) sharpens the problem: market allocation is \emph{structurally} biased toward uses where value is known in advance (routine commercial applications) and away from uses where value is uncertain but potentially transformative (research, exploration, novel inquiry).

In practice, neither markets nor governments currently allocate the question budget. Platforms do. A handful of firms---OpenAI, Anthropic, Google, xAI, Meta---that control the model layer and the inference infrastructure determine access through pricing tiers, rate limits, acceptable use policies, and content moderation rules. This is \emph{de facto} allocation by platform fiat, constrained only by competition among providers (limited, given the capital requirements) and by regulatory oversight (nascent, given the speed of deployment).

Platform allocation has the virtue of speed: firms can deploy and iterate access policies faster than legislatures can draft statutes. But it concentrates the allocation decision in entities whose objective function is shareholder value, not social welfare. The question of which questions humanity gets to ask is decided, in effect, by the terms of service of three or four companies.

The third model is direct public investment. Just as governments fund research through NSF and NIH because markets underinvest in basic knowledge \citep{arrow1962economic}, governments could fund public AI infrastructure for domains where social value exceeds private value.

The precedent is the electrical grid and the telephone network: both were initially controlled by private monopolies, both were subjected to common-carrier regulation (universal service obligations, rate regulation, non-discriminatory access), and both generated enormous positive externalities once access was democratized. The question is whether AI compute---specifically, access to tokens---should follow the same regulatory trajectory.

The choice among these models---or their combination---depends on empirical facts about transaction costs that vary across the stack. At the chip layer, where natural monopoly conditions prevail, utility-style regulation is most appropriate. At the token layer, where competition among model providers is feasible, market regulation suffices with corrections for the failures identified by Arrow and Goodhart: antitrust enforcement to prevent concentration, transparency requirements for benchmarks (to mitigate Goodhart distortion), and information disclosure rules. At the question layer, where Arrow's paradox is most acute and where the gap between social and private value is widest, some form of public allocation is likely necessary.

The Coasean insight is that no single regulatory model fits the entire stack. The optimal intervention depends on where you stand in the value chain. What is clear from the arithmetic is that the allocation question is not abstract: with 2,200 questions per person per day at the 2028 upper bound, the question budget is large enough to matter but finite enough to require choice. The choice between a medical diagnosis and a disposable video is not a technical question. It is a political one---and it should be made with the numbers in hand.

The allocation problem has a temporal dimension that current debate neglects. The question budget is distributed not only across uses at a point in time but across generations over time. The standard economic tool for intertemporal allocation is discounting: a token consumed today is valued more highly than a token consumed tomorrow, at a rate $r$ reflecting time preference and the opportunity cost of capital. But as \citet{price1993time} argues, discounting embeds an ethical judgment that is rarely examined: it systematically devalues the welfare of future persons, not because their needs are less real, but because they are further away in time. At a discount rate of 5\%, the welfare of a generation 50 years hence is weighted at $(1.05)^{-50} \approx 0.087$---less than a tenth of the present generation's weight. A question that would cure a disease in 2075 is valued at a tenth of the same question asked for commercial convenience today.

The difficulty is that the inputs to the token economy are partly exhaustible. The energy and materials committed to AI infrastructure today---copper mined, silicon refined, power plants built---are irreversibly allocated. A megawatt-hour consumed generating ephemeral tokens in 2025 cannot be consumed answering a scientific question in 2050. The efficiency gap (Section~\ref{sec:physics}) offers some relief: if energy per token falls by orders of magnitude, future generations inherit a more capable infrastructure per unit of energy. But the physical materials---copper, rare earths, water for cooling---do not benefit from algorithmic improvement. Every ton of copper drawn into a data center today is a ton unavailable for future infrastructure, whether AI or otherwise.

If the question budget is a finite physical resource, its intertemporal allocation raises the same questions that arise for any exhaustible resource: at what rate should the present generation consume computational capacity, and what obligation, if any, does it bear to preserve capacity for successors whose questions it cannot anticipate? ``Live as though you'll die tomorrow, but farm as though you'll live forever'' \citep{marsden1998tomorrow}: the token economy is governed by the first imperative, but sustainability demands the second.

The token economy is a physical economy. Tokens are units of information, and information processing is constrained by thermodynamics, channel capacity, and finite energy supply. The efficiency gap between current practice and the Landauer floor leaves room for improvement, but the floor is absolute.

The balance sheet shows the AI token economy expanding from roughly 125 tokens per person per day in mid-2024 to a projected 225,000 under aggressive infrastructure assumptions. The question budget reframes AI policy as resource allocation: finite computational capacity must be distributed across science, commerce, governance, and private consumption.

But physical expansion does not dissolve epistemic constraint. A thousand-fold increase in token throughput multiplies the rate at which answers can be generated; it does not improve the quality of the questions that elicit them, reduce the structural uncertainty in which decisions must be made, or close the gap between proxy metrics and the objectives they approximate.

Entropy reduction is not direction (Section~\ref{sec:agency}): a system may process vast quantities of information without clarifying which action improves outcomes. Optimization against imperfect proxies introduces structural distortion that grows with optimization pressure, not despite it (Section~\ref{sec:measurement}). No allocation mechanism---market, platform, or state---can fully substitute for disciplined judgment in the selection and sequencing of inquiry.

The token economy can amplify inquiry. It cannot determine which inquiries matter. That choice remains human.

\bibliography{ref}

\begin{thebibliography}{65}
\providecommand{\natexlab}[1]{#1}
\providecommand{\url}[1]{\texttt{#1}}
\expandafter\ifx\csname urlstyle\endcsname\relax
  \providecommand{\doi}[1]{doi: #1}\else
  \providecommand{\doi}{doi: \begingroup \urlstyle{rm}\Url}\fi

\bibitem[Acemoglu and Restrepo(2019)]{acemoglu2019automation}
Daron Acemoglu and Pascual Restrepo.
\newblock Automation and new tasks: How technology displaces and reinstates labor.
\newblock \emph{Journal of Economic Perspectives}, 33\penalty0 (2):\penalty0 3--30, 2019.

\bibitem[{Alphabet Inc.}(2025)]{alphabet2025q2}
{Alphabet Inc.}
\newblock Alphabet second quarter 2025 results.
\newblock \url{https://blog.google/inside-google/}, 2025.
\newblock Reported 980 trillion tokens processed monthly by {Google AI} models.

\bibitem[Altman(2024)]{altman2024openai}
Sam Altman.
\newblock Statement on {OpenAI} daily usage.
\newblock Post on X, February 2024, 2024.
\newblock Reported approximately 100 billion words generated per day.

\bibitem[Amodei et~al.(2016)Amodei, Olah, Steinhardt, Christiano, Schulman, and Man{\'e}]{amodei2016faulty}
Daron Amodei, Chris Olah, Jacob Steinhardt, Paul Christiano, John Schulman, and Dan Man{\'e}.
\newblock Concrete problems in {AI} safety.
\newblock \emph{arXiv preprint arXiv:1606.06565}, 2016.

\bibitem[Appenzeller(2024)]{appenzeller2024llmflation}
Guido Appenzeller.
\newblock {LLMflation}: {LLM} inference cost.
\newblock Andreessen Horowitz, November 2024.
\newblock URL \url{https://a16z.com/llmflation-llm-inference-cost/}.

\bibitem[Arrow(1962)]{arrow1962economic}
Kenneth~J. Arrow.
\newblock Economic welfare and the allocation of resources for invention.
\newblock In \emph{The Rate and Direction of Inventive Activity: Economic and Social Factors}, pages 609--626. Princeton University Press, 1962.

\bibitem[Bekenstein(1981)]{bekenstein1981universal}
Jacob~D. Bekenstein.
\newblock Universal upper bound on the entropy-to-energy ratio for bounded systems.
\newblock \emph{Physical Review D}, 23\penalty0 (2):\penalty0 287--298, 1981.

\bibitem[Bennett(2003)]{bennett2003notes}
Charles~H. Bennett.
\newblock Notes on {L}andauer's principle, reversible computation, and {M}axwell's demon.
\newblock \emph{Studies in History and Philosophy of Science Part B: Studies in History and Philosophy of Modern Physics}, 34\penalty0 (3):\penalty0 501--510, 2003.

\bibitem[Clark and Amodei(2016)]{clark2016faulty}
Jack Clark and Dario Amodei.
\newblock Faulty reward functions in the wild.
\newblock \url{https://openai.com/index/faulty-reward-functions/}, 2016.

\bibitem[Coase(1937)]{coase1937nature}
Ronald~H. Coase.
\newblock The nature of the firm.
\newblock \emph{Economica}, 4\penalty0 (16):\penalty0 386--405, 1937.

\bibitem[Coase(1960)]{coase1960problem}
Ronald~H. Coase.
\newblock The problem of social cost.
\newblock \emph{Journal of Law and Economics}, 3:\penalty0 1--44, 1960.

\bibitem[Coase(1972)]{coase1972durability}
Ronald~H. Coase.
\newblock Durability and monopoly.
\newblock \emph{Journal of Law and Economics}, 15\penalty0 (1):\penalty0 143--149, 1972.

\bibitem[Cook(1971)]{cook1971complexity}
Stephen~A. Cook.
\newblock The complexity of theorem-proving procedures.
\newblock In \emph{Proceedings of the Third Annual {ACM} Symposium on Theory of Computing}, pages 151--158, 1971.

\bibitem[Cox(1946)]{cox1946probability}
Richard~T. Cox.
\newblock Probability, frequency, and reasonable expectation.
\newblock \emph{American Journal of Physics}, 14\penalty0 (1):\penalty0 1--10, 1946.

\bibitem[Cox(1961)]{cox1961algebra}
Richard~T. Cox.
\newblock \emph{The Algebra of Probable Inference}.
\newblock Johns Hopkins Press, Baltimore, 1961.

\bibitem[Cox(1979)]{cox1979inference}
Richard~T. Cox.
\newblock Of inference and inquiry: An essay in inductive logic.
\newblock In Raphael~D. Levine and Myron Tribus, editors, \emph{The Maximum Entropy Formalism}, pages 119--167. MIT Press, 1979.

\bibitem[Dean(2009)]{dean2009numbers}
Jeff Dean.
\newblock Designs, lessons and advice from building large distributed systems.
\newblock Keynote, {LADIS} workshop, 2009.
\newblock Including ``Numbers Everyone Should Know''.

\bibitem[{DeepSeek-AI}(2024)]{deepseek2024v3}
{DeepSeek-AI}.
\newblock {DeepSeek-V3} technical report.
\newblock arXiv:2412.19437, 2024.

\bibitem[{Epoch AI}(2025)]{epochai2025aitrends}
{Epoch AI}.
\newblock Key trends and figures in machine learning.
\newblock \url{https://epoch.ai/trends}, 2025.

\bibitem[Giustra(2025)]{giustra2025copper}
Frank Giustra.
\newblock Is copper at the start of the next supercycle?
\newblock \url{https://frankgiustra.com/posts/is-copper-at-the-start-of-the-next-supercycle/}, 2025.

\bibitem[{Goldman Sachs Research}(2024)]{goldmansachs2024copper}
{Goldman Sachs Research}.
\newblock Generational growth: {AI}, data centers and the coming {US} power demand surge.
\newblock \url{https://www.goldmansachs.com/insights/articles/AI-poised-to-drive-160-increase-in-power-demand}, 2024.

\bibitem[Goodhart(1984)]{goodhart1984monetary}
Charles A.~E. Goodhart.
\newblock \emph{Monetary Theory and Practice: The {UK} Experience}.
\newblock Macmillan, London, 1984.

\bibitem[Grantham(2011)]{grantham2011resources}
Jeremy Grantham.
\newblock Time to wake up: Days of abundant resources and falling prices are over forever.
\newblock GMO Quarterly Letter, April 2011, 2011.

\bibitem[Heisenberg(1927)]{heisenberg1927ueber}
Werner Heisenberg.
\newblock {\"U}ber den anschaulichen {I}nhalt der quantentheoretischen {K}inematik und {M}echanik.
\newblock \emph{Zeitschrift f{\"u}r Physik}, 43\penalty0 (3--4):\penalty0 172--198, 1927.

\bibitem[{International Energy Agency}(2024)]{iea2024electricity}
{International Energy Agency}.
\newblock Electricity 2024: Analysis and forecast to 2026.
\newblock \url{https://www.iea.org/reports/electricity-2024}, 2024.

\bibitem[{International Energy Agency}(2025)]{iea2025datacenter}
{International Energy Agency}.
\newblock Energy and {AI}: Data centres, the new frontier of energy demand.
\newblock \url{https://www.iea.org/reports/energy-and-ai}, 2025.

\bibitem[Jaynes(2003)]{jaynes2003probability}
Edwin~T. Jaynes.
\newblock \emph{Probability Theory: The Logic of Science}.
\newblock Cambridge University Press, 2003.

\bibitem[Jevons(1865)]{jevons1865coal}
William~Stanley Jevons.
\newblock \emph{The Coal Question: An Inquiry Concerning the Progress of the Nation, and the Probable Exhaustion of Our Coal-Mines}.
\newblock Macmillan, London, 1865.

\bibitem[Karp(1972)]{karp1972reducibility}
Richard~M. Karp.
\newblock Reducibility among combinatorial problems.
\newblock In Raymond~E. Miller and James~W. Thatcher, editors, \emph{Complexity of Computer Computations}, pages 85--103. Plenum Press, New York, 1972.

\bibitem[Keynes(1930)]{keynes1930possibilities}
John~Maynard Keynes.
\newblock Economic possibilities for our grandchildren.
\newblock In \emph{Essays in Persuasion}, pages 358--373. W.~W.~Norton, New York, 1930.

\bibitem[Landauer(1961)]{landauer1961irreversibility}
Rolf Landauer.
\newblock Irreversibility and heat generation in the computing process.
\newblock \emph{IBM Journal of Research and Development}, 5\penalty0 (3):\penalty0 183--191, 1961.

\bibitem[Levitt and Dubner(2009)]{levitt2009superfreakonomics}
Steven~D. Levitt and Stephen~J. Dubner.
\newblock \emph{Superfreakonomics: Global Cooling, Patriotic Prostitutes, and Why Suicide Bombers Should Buy Life Insurance}.
\newblock William Morrow, New York, 2009.

\bibitem[Lloyd(2000)]{lloyd2000ultimate}
Seth Lloyd.
\newblock Ultimate physical limits to computation.
\newblock \emph{Nature}, 406\penalty0 (6799):\penalty0 1047--1054, 2000.

\bibitem[Luccioni et~al.(2024)Luccioni, Jernite, and Strubell]{luccioni2024power}
Alexandra~Sasha Luccioni, Yacine Jernite, and Emma Strubell.
\newblock Power hungry processing: Watts driving the cost of {AI} deployment?
\newblock In \emph{Proceedings of the 2024 ACM Conference on Fairness, Accountability, and Transparency}, pages 85--99, 2024.

\bibitem[MacKay(2003)]{mackay2003information}
David J.~C. MacKay.
\newblock \emph{Information Theory, Inference, and Learning Algorithms}.
\newblock Cambridge University Press, 2003.

\bibitem[MacKay(2009)]{mackay2009sustainable}
David J.~C. MacKay.
\newblock \emph{Sustainable Energy---Without the Hot Air}.
\newblock UIT Cambridge, 2009.

\bibitem[Marsden(1998)]{marsden1998tomorrow}
John Marsden.
\newblock Remarks on sustainable land management, 1998.
\newblock Widely quoted in Australian agricultural literature.

\bibitem[Meadows et~al.(1972)Meadows, Meadows, Randers, and Behrens~III]{meadows1972limits}
Donella~H. Meadows, Dennis~L. Meadows, J{\o}rgen Randers, and William~W. Behrens~III.
\newblock \emph{The Limits to Growth}.
\newblock Universe Books, New York, 1972.

\bibitem[Mehl et~al.(2007)Mehl, Vazire, Ram{\'\i}rez-Esparza, Slatcher, and Pennebaker]{mehl2007women}
Matthias~R. Mehl, Simine Vazire, Nair{\'a}n Ram{\'\i}rez-Esparza, Richard~B. Slatcher, and James~W. Pennebaker.
\newblock Are women really more talkative than men?
\newblock \emph{Science}, 317\penalty0 (5834):\penalty0 82, 2007.

\bibitem[Moore(1965)]{moore1965cramming}
Gordon~E. Moore.
\newblock Cramming more components onto integrated circuits.
\newblock \emph{Electronics}, 38\penalty0 (8):\penalty0 114--117, 1965.

\bibitem[{Nebius AI}(2025)]{nebius2025economics}
{Nebius AI}.
\newblock The economics of {AI} clusters.
\newblock Nebius whitepaper, 2025.

\bibitem[{NVIDIA}(2024)]{nvidia2024blackwell}
{NVIDIA}.
\newblock {NVIDIA Blackwell} platform arrives to power a new era of computing.
\newblock \url{https://nvidianews.nvidia.com/news/nvidia-blackwell-platform-arrives-to-power-a-new-era-of-computing}, 2024.

\bibitem[{OpenRouter} and {Andreessen Horowitz}(2025)]{openrouter2025stateofai}
{OpenRouter} and {Andreessen Horowitz}.
\newblock State of {AI} 2025: An empirical 100 trillion token study.
\newblock \url{https://a16z.com/state-of-ai/}, 2025.

\bibitem[Patterson et~al.(2021)Patterson, Gonzalez, Le, Liang, Munguia, Rothchild, So, Texier, and Dean]{patterson2021carbon}
David Patterson, Joseph Gonzalez, Quoc Le, Chen Liang, Lluis-Miquel Munguia, Daniel Rothchild, David So, Maud Texier, and Jeff Dean.
\newblock Carbon emissions and large neural network training.
\newblock \emph{arXiv preprint arXiv:2104.10350}, 2021.

\bibitem[Pierrehumbert(2009)]{pierrehumbert2009levitt}
Raymond~T. Pierrehumbert.
\newblock An open letter to {Steve Levitt}.
\newblock RealClimate, October 2009.
\newblock URL \url{https://www.realclimate.org/index.php/archives/2009/10/an-open-letter-to-steve-levitt/}.

\bibitem[Price(1993)]{price1993time}
Colin Price.
\newblock \emph{Time, Discounting and Value}.
\newblock Blackwell, Oxford, 1993.

\bibitem[Prigogine(1978)]{prigogine1978time}
Ilya Prigogine.
\newblock Time, structure, and fluctuations.
\newblock \emph{Science}, 201\penalty0 (4358):\penalty0 777--785, 1978.

\bibitem[Russell(1935)]{russell1935idleness}
Bertrand Russell.
\newblock \emph{In Praise of Idleness and Other Essays}.
\newblock George Allen \& Unwin, London, 1935.

\bibitem[Schwager(1989)]{schwager1989wizards}
Jack~D. Schwager.
\newblock \emph{Market Wizards: Interviews with Top Traders}.
\newblock New York Institute of Finance, New York, 1989.

\bibitem[Shannon(1948)]{shannon1948mathematical}
Claude~E. Shannon.
\newblock A mathematical theory of communication.
\newblock \emph{Bell System Technical Journal}, 27\penalty0 (3):\penalty0 379--423, 1948.

\bibitem[Shannon(1951)]{shannon1951prediction}
Claude~E. Shannon.
\newblock Prediction and entropy of printed {E}nglish.
\newblock \emph{Bell System Technical Journal}, 30\penalty0 (1):\penalty0 50--64, 1951.

\bibitem[Shumailov et~al.(2024)Shumailov, Shumaylov, Zhao, Papernot, Anderson, and Gal]{shumailov2024collapse}
Ilia Shumailov, Zakhar Shumaylov, Yiren Zhao, Nicolas Papernot, Ross Anderson, and Yarin Gal.
\newblock {AI} models collapse when trained on recursively generated data.
\newblock \emph{Nature}, 631\penalty0 (8022):\penalty0 755--759, 2024.

\bibitem[Simon(1980)]{simon1980resources}
Julian~L. Simon.
\newblock Resources, population, environment: An oversupply of false bad news.
\newblock \emph{Science}, 208\penalty0 (4451):\penalty0 1431--1437, 1980.

\bibitem[Simon(1981)]{simon1981ultimate}
Julian~L. Simon.
\newblock \emph{The Ultimate Resource}.
\newblock Princeton University Press, 1981.

\bibitem[Sivathanu et~al.(2024)Sivathanu, Zhao, and Reddi]{sivathanu2024revisiting}
Muthian Sivathanu, Yijia Zhao, and Vijay~Janapa Reddi.
\newblock Revisiting reliability in large-scale machine learning research clusters.
\newblock \emph{arXiv preprint arXiv:2410.21680}, 2024.

\bibitem[Soros(1987)]{soros1987alchemy}
George Soros.
\newblock \emph{The Alchemy of Finance: Reading the Mind of the Market}.
\newblock Simon and Schuster, New York, 1987.

\bibitem[Strubell et~al.(2019)Strubell, Ganesh, and McCallum]{strubell2019energy}
Emma Strubell, Ananya Ganesh, and Andrew McCallum.
\newblock Energy and policy considerations for deep learning in {NLP}.
\newblock In \emph{Proceedings of the 57th Annual Meeting of the Association for Computational Linguistics}, pages 3645--3650, 2019.

\bibitem[{TSMC}(2024)]{tsmc2024arizona}
{TSMC}.
\newblock {TSMC} arizona: Third fab announcement and {CHIPS Act} funding.
\newblock \url{https://pr.tsmc.com/english/news/3122}, 2024.

\bibitem[Turing(1936)]{turing1936computable}
Alan~M. Turing.
\newblock On computable numbers, with an application to the {E}ntscheidungsproblem.
\newblock \emph{Proceedings of the London Mathematical Society}, s2-42\penalty0 (1):\penalty0 230--265, 1936.

\bibitem[{U.S. Energy Information Administration}(2024)]{eia2024electricity}
{U.S. Energy Information Administration}.
\newblock Electric power annual 2023.
\newblock \url{https://www.eia.gov/electricity/annual/}, 2024.

\bibitem[von Neumann(1958)]{vonneumann1958computer}
John von Neumann.
\newblock \emph{The Computer and the Brain}.
\newblock Yale University Press, New Haven, 1958.

\bibitem[von Neumann(1966)]{vonneumann1966selfreproducing}
John von Neumann.
\newblock \emph{Theory of Self-Reproducing Automata}.
\newblock University of Illinois Press, Urbana, 1966.
\newblock Edited and completed by Arthur W. Burks.

\bibitem[von Neumann and Morgenstern(1944)]{vonneumann1944games}
John von Neumann and Oskar Morgenstern.
\newblock \emph{Theory of Games and Economic Behavior}.
\newblock Princeton University Press, 1944.

\bibitem[Wiener(1948)]{wiener1948cybernetics}
Norbert Wiener.
\newblock \emph{Cybernetics: Or Control and Communication in the Animal and the Machine}.
\newblock MIT Press, Cambridge, MA, 1948.

\bibitem[Wiener(1950)]{wiener1950human}
Norbert Wiener.
\newblock \emph{The Human Use of Human Beings: Cybernetics and Society}.
\newblock Houghton Mifflin, Boston, 1950.

\end{thebibliography}
\end{document}